# UFCORIN: A Fully Automated Predictor of Solar Flares in GOES X-Ray Flux.

Takayuki Muranushi,[1] Takuya Shibayama,[2] Yuko Hada Muranushi,[3] Hiroaki Isobe,[4,5] Shigeru Nemoto,[5,6] Kenji Komazaki,[6] Kazunari Shibata,[3,5]

———

[1] RIKEN Advanced Institute for Computational Science, 7-1-26, Minatojima-minami-machi, Chuo-ku, Kobe, Hyogo, 650-0047, Japan

[2] Solar-Terrestrial Environment Laboratory, Nagoya University, Chikusa-ku, Nagoya, 464-8601, Japan

[3] Kwasan and Hida Observatories, Kyoto University, Yamashina-ku, Kyoto 607-8471, Japan

[4] Graduate School of Advanced Integrated Studies for Human Survivability, Kyoto University Yoshida Nakaadachi-cho, Sakyo-ku, Kyoto 606-8306, Japan

[5] Unit of Synergetic Studies for Space, Kyoto University, Kitashirakawa

D R A F T          October 31, 2015, 3:00pm          D R A F T

**Abstract.** We have developed UFCORIN, a platform for studying and automating space weather prediction. Using our system we have tested 6,160 different combinations of SDO/HMI data as input data, and simulated the prediction of GOES X-ray flux for 2 years (2011-2012) with one-hour cadence. We have found that direct comparison of the true skill statistics (TSS) from small cross-validation sets is ill-posed, and used the standard scores ($z$) of the TSS to compare the performance of the various prediction strategies. The $z$ of a strategy is a stochastic variable of the stochastically-chosen cross-validation dataset, and the $z$ for the three strategies best at predicting X, $\geq$M and $\geq$C class flares are better than the average $z$ of the 6,160 strategies by $2.3\sigma$, $2.1\sigma$, $3.8\sigma$ confidence levels, respectively. The best three TSS values were $0.75\pm0.07$, $0.48\pm0.02$, and $0.56\pm0.04$, respectively.

Oiwake-cho, Sakyo-ku, Kyoto 606-8502,

Japan

[6]BroadBand Tower, Inc. 1-3-2

Uchisaiwai-cho, Chiyoda-ku, Tokyo

100-0011, Japan

## 1. Introduction

Prediction of the onset of solar flares and associated eruptions (e.g. *Shibata and Magara* [2011]) is one of the most important goals of space weather studies (e.g. *Baker* [2004]; *Schwenn* [2006]). Numerous research groups have been working to establish flare prediction based on dynamical models. Several numerical models have been developed and regularly used to predict the solar wind using the observations of the photosphere (e.g., SUSANOO; *Miyoshi and Kataoka* [2008]; *Shiota et al.* [2014], http://st4a.stelab.nagoya-u.ac.jp/susanoo/index.html and WSA-ENLIL; *Odstrcil et al.* [2004], http://www.swpc.noaa.gov/products/wsa-enlil-solar-wind-prediction ). However, these solar wind predictions suffer from poorly-know initial conditions, because calculating the pre-flare coronal magnetic fields from observed magnetic field at the photosphere and simulating the flare onset are still challenging tasks [*Bamba et al.*, 2013; *Savani et al.*, 2015].

A viable alternative is to develop an empirical algorithm that calculates the probability of flare occurrence from observational data, using the statistics of flare onset in the past. Since flares are explosive release of magnetic energy stored in the corona, most of previous studies in this line used photospheric magnetograms or white light images. There are various approaches to flare prediction: discriminant analysis of the magnetic parameters obtained from vector magnetic fields [*Leka and Barnes*, 2003], superposed epoch analysis [*Mason and Hoeksema*, 2010], Bayesian statistics [*Wheatland*, 2005], statistical analysis based on McIntosh sunspot classification [*Bloomfield et al.*, 2012], etc. Recently, several flare prediction algorithms have been developed using machine learning techniques: support vector machines [*Li et al.*, 2007; *Bobra and Couvidat*, 2014], ordinal logistic regression [*Song et al.*, 2009; *Yuan et al.*, 2010], neural networks [*Colak and Qahwaji*, 2009; *Yu et al.*, 2009; *Ahmed et al.*, 2013]. Such techniques can be applied not only to the flare prediction but also to the prediction of other important parameters of space weather, such as wind velocity, magnetic field and density of solar wind, flux of solar energetic particles, and various indices of geomagnetic disturbance.

We have developed UFCORIN (Universal Forecast Constructor by Optimized Regression of INputs), a new space-weather prediction engine. As the name suggests, UFCORIN is designed as a generic time-series predictor, which can be configured to predict any time series variable from an arbitrary set of input time series. Therefore, we can use UFCORIN to predict various target parameters by choosing appropriate input time series from observations. This design of UFCORIN allows the users to flexibly change the input and target parameters when more advanced observational data become available, or when needs for new prediction targets arise.

As the first step of the development, we present the result of the flare prediction by UFCORIN. The prediction target parameter is the maximum of the GOES soft X-ray flux (1-8 Å) in the next 24 hours, and the input parameters are the wavelet power of the full disk line-of-sight magnetogram obtained by the Helioseismic and Magnetic Imager (HMI; *Scherrer et al.* [2012]) on board the Solar Dynamic Observatory (SDO; *Schou et al.* [2012]). Following *Bloomfield et al.* [2012], we evaluate the performance of the flare prediction engine UFCORIN using the True Skill Statistic (TSS).

## 2. Overview of The Method

Figure 1 illustrates the prediction pipeline of our space weather forecast engine, which predicts the GOES X-ray flux in the following procedure.

1. Read a prediction strategy containing a selected list of the time series data and other prediction configurations. Load the specified input and output time series from the open-access solar observation archive data.

2. Separate the data into a training set $t_j$ and a test set $t_k$ in order to perform cross validation.

3. Construct optimized regression predictor of the model function $y(\vec{x})$ from the training set data.

4. Predict $y'_k$ for the test data, by applying the test data input vectors $\vec{x}(t_k)$ to the learned prediction function $y(\vec{x})$.

5. Predict X,$\geq$M,$\geq$C class flares by comparing the $y'_k$ to thresholds. The thresholds are selected to maximize the TSS, using the method of *Bloomfield et al.* [2012].

6. Compare the prediction and observation results, and generate the contigency tables.

7. Calculate true skill statistics (TSS).

In the study covered by this paper, we constructed input vectors $\vec{x}(t_i)$ by applying wavelet transformation to solar images (c.f. §3.2), and output data $y(t_i)$ from the GOES X-ray flux data (c.f. §3.3). The details of experiments are described in §4, including how we conducted cross validation, prediction, TSS calculation , and noise estimation. Technical details of UFCORIN are described in the Appendices of this paper (c.f. A1 - A3).

A contingency table is a 2 × 2 table (e.g. Table 1) that shows the numbers of true positive (TP), true negative (TN), false positive (FP), and false negative (FN) cases. Using a contingency table, the TSS is calculated as follows:

$$TSS = \frac{TP}{TP+FN} - \frac{FP}{FP+TN}. \tag{1}$$

There are two reasons why we use TSS to measure the prediction quality. First, TSS is one of the skill indicators that equals to 0 when the predictor has no knowledge of the event, and equals to 1 when the prediction is perfect (i.e. no false positives nor false negatives.) Second, TSS is one of the indicators that are not affected by the change of ratio between positive and negative events, provided that the events are sampled from a same population. Because of these two reasons, previous studies [*Bloomfield et al.*, 2012; *Bobra and Couvidat*, 2014] uses TSS as the preferred indicator for measuring the performance of predictions, and we also use TSS.

Optionally, UFCORIN can be executed in automated optimization mode. In automated optimization mode, UFCORIN repeats the steps 3.-7., changing the machine learning parameters. UFCORIN automatically finds the set of the parameters that maximizes the results. The maximization target can be the TSS for a specific class of flares, or any weighted average of the TSS values for all classes.

## 3. Preparation of the Data

### 3.1. Design of Time Series Data Handling in UFCORIN

In order to construct flare prediction models, we have used SDO/HMI data as the input, and GOES X-ray flux as the output. In essence, our motivation is to predict the solar X-ray flux from magnetic field images of the Sun. We used data from the beginning of year 2011 to the end of year 2012, UTC, with a cadence of one hour. We adopt the one-hour cadence throughout this study because it is sufficiently smaller than the evolution timescale of the active regions (ARs). Flare-triggering magnetic field structures are observed hours before the flare events [*Kusano et al.*, 2012; *Bamba et al.*, 2013], and brightening in X-ray lasts hours after the onset of the flare. One-hour cadence is small enough to resolve these evolutions. On the other hand, the one-hour cadence is long enough so that one prediction experiment for the two-year period ends in a few minutes, allowing us to carry out the survey over prediction strategies in a plausible amount of time.

UFCORIN is designed to accept fixed number of time series data and output another time series. Because of this design, all the input and output data have to be variables that are globally defined at any given moment of the Sun. In the current mode of operation, we do not use variables that are defined per sunspot, or per active region.

This makes UFCORIN easier to operate in an automated manner. UFCORIN does not require human intervention to set up sunspot or active region detection. This means that UFCORIN can provide quicker flare predictions since it does not wait for sunspot or active region detection results. This also means that UFCORIN is free from possible biases of such detection processes. Instead, UFCORIN is able to learn how to detect features. In UFCORIN the feature detection is integrated with other parts of the prediction pipeline, and the whole system is optimized as a predictor. This creates more opportunities for improvement of prediction.

Nevertheless, it is still possible to use per-sunspot or per-active-region data. Since there can be multiple sunspots and active regions on the Sun at any given moment, such data are multi-valued functions of time. From such multi-valued functions, we may create single-valued functions of time that are suitable as input data for UFCORIN. For example, we can take the sum, or the average over feature values defined per sunspot. For the time points where there is no sunspots, we can use a default value of, say, 0.

### 3.2. Input Data

We used 45-second SDO/HMI line-of-sight magnetic field strength map `hmi.M_45s` to construct the input data. We sampled the first data of every one-hour period, to construct an image time series with cadence of one hour for the two-year period of 2011-2012. We have preprocessed the images to construct various scalar time series data that reflect different features of the Sun.

We construct input data by integrating the absolute vaule of the magnetic field strength found in the HMI images. We also construct input data by applying wavelet analysis to the HMI images. Wavelet transformations are used to quantify the complexity across the scales, and have been applied to flare productivity prediction studies [*Hewett et al.*, 2008]. In addition, we import data from GOES X-ray lightcurve for past time intervals, by taking its average, squared average, and maximum over 24, 48 and 72 hours period in the past. All these time series are also calculated at one-hour cadence.

Our method of preprocessing SDO/HMI images is as follows. First, we downloaded the original FITS images from the SDO/HMI website using the script program (http://jsoc.stanford.edu/ajax/exportfile.csh) provided by JSOC (Joint Science Operations Center). The size of the original FITS images is $4096 \times 4096$. The Sun is present as a large disk in the images. However, pixels close to the rim of the disk are either off the solar surface or too noisy, so we remove such rim pixels.

In order to remove the rim, we set all the pixels to zero whose distance from the image center is greater than 1792. Given that the original SDO/HMI images are sampled at 0.5 arcsec/pixel, one pixel corresponds to approximately $3.63 \times 10^7$ cm at 1 AU. Given that the solar radius is $6.96 \times 10^{10}$ cm ($4.65 \times 10^{-3}$ AU), our rim noise filter corresponds to approximately 93% of the solar radius, or 69 degrees in terms of the angle from the solar front. Next, we scaled each of the original $4096 \times 4096$ image to size $1024 \times 1024$ by averaging every $4 \times 4$ pixels. Then, we applied a two-dimensional (2D) discrete wavelet transformation to the images.

There are many variations of 2D wavelet transformations. Table 2 lists the wavelet transformations we used in this study, together with their labels. The details of the wavelet transformation are described in e.g. *Arai* [2010]. We provide here a brief introduction to wavelets, and present Figure 2, which visualizes some of the wavelet basis functions.

First, as in Figure 2, there are two kinds of 2D wavelet transformations, standard (S) and non-standard (N). They are distinguished by that the order the x- and y-direction wavelet transformations are interleaved. Also, there are many wavelet basis functions with different waveforms. Among them, we use Haar basis and the $\beta$-spline basis in this study. We choose the two bases because they both have simple waveform with Haar basis being discontinuous and $\beta$-spline basis being smooth. Their waveforms are shown in Figure 2. By choosing a wavelet basis and either standard or non-standard transformation, we can construct many different 2D wavelet transformations.

In order to use wavelet features in UFCORIN, we need to create scalar functions of time. Now, wavelet-transformed image consists of multiple rectangular regions, indicated as green rectangles in Figure 2. Each of these regions represents different horizontal/vertical scales. Therefore we calculate the integral and square-integral of the amplitude for every region, and use their time series as input data.

Figure 3 shows the effect of the wavelet transformation on an SDO/HMI image. Since we have reduced the image resolution to $1024^2$ before the wavelet transformation, there are $\log_2 1024 + 1 = 11$ distinct regions per edge of the wavelet space. Therefore, number of distinct regions produced by standard and non-standard 2D wavelat transformation are 121 and 31, respectively. Since we calculate the integral and the square-integral for each region, the number of input data is twice the number of the regions. Therefore, we obtain 242 and 62 input data values from a single wavelet-transformed image, respectively. For prediction, we only use these timline data.

Thus, the number of input time series introduced via wavelet transformation is 304 multiplied by the number of different wavelet basis functions we use. Although UFCORIN can handle input data sets of arbitrary size, larger input data sets result in more execution time for prediction. A rough estimate is that a single learning/prediction cycle completes in a few minutes, for 2-years worth of data, using a single node of our current system (c.f. §A3). When in automated optimization mode, UFCORIN repeats the prediction using CMA-ES algorithm [*Hansen and Ostermeier*, 1996, 2001] until she concludes that the TSS is optimized. The convergence criteria for optimization can be specified in the strategy file. In our study, we stop optimization when the TSS improvement of the last step of CMA-ES was smaller than $10^{-3}$. With this choice, the optimization takes 100 to 500 minutes.

### 3.3. Output Data

For the output data, we use the GOES soft X-ray (1-8 Å) flux data published at *Space Weather Prediction Center Website* [2014a], since the 1-8 Å flux of the Sun is the widely-used measure of the solar flare intensity. Again, we use the data from the beginning of 2011 to the end of 2012. In order to construct one-hour-cadence time series data from the GOES data, we first retrieved the GOES XRS (X-ray sensor) two-second-cadence data, and then binned them into one-hour cadence by taking the maximum of every timebin. Note that the standard definition of a GOES X-ray flare magnitude is based on 1-minute averages of the higher cadence XRS data. The difference have caused less than one percent of the flare events to be classified differently by our method and by the standard definition.

The total solar X-ray flux is a variable integrated over entire solar image. Hence it is suitable for continuous prediction. Yet, the X-ray flux curve exhibits peaks from individual flares, allowing for flare predictions. However, predicting the exact soft X-ray light curve of the individual flares is extremely difficult. Therefore, instead of attempting to predict individual events, we treat solar flares as stochastic events and introduce statistical variables as prediction targets.

Such variables must be tractable for prediction, and at the same time must possess applicational utility. We have chosen to predict the *future maximum* of the solar X-ray flux for a given timespan. Here, for a given time series $y_0(t')$, its future maximum of span $T$ is denoted as $y_{\max,T}(t)$, and defined as follows:

$$y_{\max,T}(t) = \max_{t<t'<t+T} y_0(t'). \tag{2}$$

Predicting the future maximum is equivalent to predicting the largest flare event that will occur at some time in the given timespan. For example, the prediction that the 24-hour future maximum of solar X-ray flux will be greater than $10^{-4}\text{Wm}^{-2}$ means that there will be at least one X-class flare events within next 24 hours. Note that this is different from predicting that an X-class flare will take place *exactly* after 24 hours from the time of the forecast.

In this study, we predict three different classes of timebins (events defiend at every hour). The three classes are labeled X, $\geq$M, and $\geq$C, respectively.

- A positive timebin $t$ of class X is where $y_{\text{max,24hr}}(t) \geq 10^{-4}\text{Wm}^{-2}$.
- A positive timebin $t$ of class $\geq$M is where $y_{\text{max,24hr}}(t) \geq 10^{-5}\text{Wm}^{-2}$.
- A positive timebin $t$ of class $\geq$C is where $y_{\text{max,24hr}}(t) \geq 10^{-6}\text{Wm}^{-2}$.

In other words, a positive timebin of $\geq$M means that there will be at least one flare larger or equal to M class in 24 hours, and a positive timebin of $\geq$C means that there will be at least one flare larger or equal to C class in 24 hours.

We adopt this classification of events because in this way the events have simple representations using $y_{\text{max},T}$. We can nevertheless predict the individual classes of flares by composing multiple predictions. For example, if UFCORIN gives a positive prediction to $\geq$C-class but a negative prediction to $\geq$M-class, it is effectively predicting an occurence of one or more C-class flares and no $\geq$M-class flares.

In Figure 4, the red points shows the predicted future maximum values $y'_{\text{max,24hr}}(t_k)$, calculated at step 4. of UFCORIN's prediction pipeline, as described in §2. Red dots are plotted at cadence of one-hour, and each dot at $t$ corresponds to one prediction of 24-hour future from the time $t$.

If the regression is perfect, the red points that corresponds to $y'_{\text{max,24hr}}(t_k)$ should exactly match the blue curve, which represents $y_{\text{max,24hr}}(t)$. In reality, regression is not perfect, and the red points do not exactly match the blue curve. Especially, note that no red points are above $10^{-4} \text{Wm}^{-2}$ line; had we used the raw values of $y'_{\text{max,24hr}}(t_k)$ for prediction, we will predict no X-class positive events.

We still predict X-class positive events, because at step 5. of the prediction pipeline we adjust the thresholds for predicting given class of events. In case of Figure 4, the threshold for X-class is $10^{-5.1919} \text{Wm}^{-2}$, indicated by the thick dashed gray line. This choice of threshold causes UFCORIN to predict X-class positive events for $5275 \leq t \leq 5312$, while X-class positive events are observed for $5265 \leq t \leq 5288$. The change of the threshold from the original definition ($10^{-4} \text{Wm}^{-2}$) has increased the numbers of both the true positives and the false positives, resulting in the increase of the TSS. Similarly, the thresholds for $\geq$M-class and $\geq$C-class events were $10^{-5.4463} \text{Wm}^{-2}$ and $10^{-5.7055} \text{Wm}^{-2}$, respectively, for this case.

We include the threshold optimization in the pipeline, because regression algorithms tend to be biased towards most frequent events, and we have to adjust for the bias. However, it is difficult to adjust the regression algorithms themselves to make accurate prediction for infrequent events.

One way to make such an adjustment is to apply a regressor post-process function $f_p$ that cancels the most-frequent-data bias, by stretching the output values of the regression function around the most-frequent data. For example, an $f_p$ may map $10^{-5.1919} \text{Wm}^{-2}$ to $10^{-4} \text{Wm}^{-2}$, $10^{-5.4463} \text{Wm}^{-2}$ to $10^{-5} \text{Wm}^{-2}$, and $10^{-5.7055} \text{Wm}^{-2}$ to $10^{-6} \text{Wm}^{-2}$. The knowledge of these three points are sufficient to make predictions for X-class, $\geq$M-class and $\geq$C-class events; we do not need the knowledge of the full form of the function $f_p$.

Thus, adjusting the thresholds can be interpreted as a convenient and practical method for making unbiased predictions.

## 4. Prediction Experiments and Results

### 4.1. Outline of the Prediction Experiment

We address the following question in the experiment: what is the input data set that achieves the best prediction outcome, as measured by the true skill statistics?

As listed in §3.2, the input data we have prepared are:

- The average, squared average, and maximum of GOES X-ray flux over past 24, 48 and 72 hours.

- The integral of the absolute value of the magnetic field strength in the HMI images.

- Wavelet features of the HMI images.

In the experiment, the past-maximum of GOES X-ray flux features and the integral of the absolute magnetic field features were always included in the prediction strategies. The average and the squared average of past GOES X-ray flux was not used in this experiment. We prepared many different prediction strategies that differ in the subset of the wavelet features they include, and measured their prediction skills.

Since features in the wavelet space span many different scales on the Sun, it is possible that some of the features do not contain useful information for prediction. Such features act as noise to the predictor, and the TSS tends to decrease when we include such noise data in the input set. This happens because the machine learning engines is prone to "learning" from the superficial correlation that may exist in the training-data portion of the noise and the output. Such learned "knowledge", of course, does not generalize to the test-data portion.

This issue cannot be solved by machine learning engines because their only source of learning is the training data. One way to reduce the false correlation effect is to provide more test data. The other way is to apply the prior knowledge to select the input data, so that noisy data are removed from the input set.

In aim to study the latter effect, we apply lower and upper limit of the scale of the wavelet features that are included in the prediction strategy, and tested how the TSS distribution change in response to the change in the input feature set. However, we must keep in mind that TSS of a single prediction might be product of mere luck, since our prediction is based on a finite number of observational data. Therefore it is important to estimate the probability distribution of the TSS. We measured the probability distribution, by repeating the cross-validation experiments on different training-test data sets. The details of experiment are as follows.

**4.2. Construction of the Strategy Set**

In this section we describe how we constructed the set of prediction strategies $S$, the subject of the survey. Each strategy $s \in S$ is characterized by its wavelet basis $w(s)$, and its lower($l$) and upper($u$) bounds of the wavenumber in the horizontal and the vertical directions: $k_{hl}(s), k_{vl}(s), k_{hu}(s),$ and $k_{vu}(s)$.

The set $S$ is constructed from its elements $s$ as follows. First, $w(s)$ is one of `haar-S`, `haar-N`, `bspl-S`, or `bspl-N`. Next, as described in §3.2, our wavelet features cover 10 different spatial scales, each scale being two times larger than its predecessor, so that $k_{hl}(s), k_{hu}(s), k_{vl}(s), k_{vu}(s) \in \{2^n | 0 \leq n \leq 9\}$.

Now, there are $_{11}C_2 = 55$ different ways of choosing a lower ($l$) and an upper ($u$) limit from the 10 scales that satisfies $0 \leq l \leq u \leq 9$. For standard wavelets, we can choose two distinct scale ranges, one for the horizontal and one for the vertical direction. For non-standard wavelets, we can choose only one scale range, since the scale for the two directions are coupled: $k_{hl}(s) = k_{vl}(s)$, $k_{hu}(s) = k_{vu}(s)$. This means that we have $55^2$ different strategies per one standard wavelet basis, and 55 per one non-standard basis.

Thus, we constructed prediction strategies that contain wavelet features only from that specific basis, for all possible choices of the lower and upper limits, for all choices of the wavelet bases. Since we have two flavors of wavelet bases (Haar and $\beta$-spline), we have

$$2 \times (55^2 + 55) = 6160 \tag{3}$$

different strategies in total.

We define conversion of the wavenumber $k(s)$ into other quantities, to simplify the interpretation of our study. Particularly, we convert $k$ to $r$, the ratio of the physical scale of the feature to the solar diameter. According to the size parameters given in §3.2, $r = 1.07/k$. Since the smallest scale in the real space corresponds to the largest in wavenumber, and vice versa, $r_{hl} = 1.07/k_{hu}$ and $r_{hu} = 1.07/k_{hl}$. The relative scales in the vertical direction, $r_{vl}$ and $r_{vu}$, are defined in a similar manner.

### 4.3. Construction of the Cross-Validation Data Sets

We created 10 different training-test data set pairs (cross-validation data, or CV data), and performed prediction. The data from two-year period are divided into weeks, and a week contains 168 timebins (a timebin is an one-hour period that corresponds to the cadence.)

The method we used to create the CV data is illustrated in Figure 5. First, we divided the two-year timeline into weekly segments, and assigned training data and test data alternately (Figure 5 [0]). Then, we construct another CV data by randomly swapping the role of the training data weeks with their successive test data weeks, each with probability of 1/2 (Figure 5 [1]). The next CV data ([2]) is the negative of the previous CV data. Thus, the pair of CV data ([1],[2]) contains every week as a training data in one set and as a test data in the other. We can generate an arbitrary number of CV data sets in this method.

Next, we apply screening to the CV data set, so that the ratio of flaring/non-flaring event is maintained in the CV set. This screening is important because without it, the test set might contain fewer hard-to-predict events than the average. In such cases the TSS of the prediction tends to be better than usual. In other words, we want to homogenize the CV data with respect to prediction difficulty of the screening.

On screening, we accept CV data only if the the number of the X-class-positive timebins is no more than 110% of that of the test set, and vice versa. In other words, we require that the test set and the training set each contain more than 47.6% ($=\frac{100}{100+110}$) of the total X-class-positive timebins in the original two-year period. We screen the CV data by the ratio of the numbers of $\geq$M-class and $\geq$C-class-positive timebins with the same criteria. We repeat the process until we obtain 10 CV data, with 5 pairs of 2-fold cross validation. Table 3 lists the numbers of X-, $\geq$M-, and $\geq$C-class-positive timebins in the training and the test set of each CV data.

Note the relation between positive timebins and flare events for given class of flares. For example, there has been 17 X-class flares in two-year period 2011-2012, according to RHESSI database (*Schwartz et al.* [2002]). This should give rise to $17 \times 24 = 408$ X-class-positive timebins, had all the flare events been independent. In reality, however, multiple X-class flares can take place in less than 24 hours. For example, three X-class flare events took place at 10:53:16, 11:06:40, and 11:36:04 on 2011-09-22. Due to such overlaps, there are only 339 X-class-positive timebins in the studied two-year period.

There are $(365 + 366) \times 24 = 17544$ timebins in the two-year period (Note that 2012 is a leap year.) Of the 731 days in this period, the first 3 days and the last 1 day are not subject of the prediction study, since we use 72-hour past to construct the input data, and 24-hour future to construct the prediction target. We also omit any timebins, when the input or output data are not available due to the lack of observational data. There are 702 timebins omitted in this way. As the result, there are 16746 timebins in the CV data. The number of timebin in the test set is approximately $8373 = 16746/2$. However, this number fluctuates as the weeks with missing data are randomly assigned to the test set and the training set.

### 4.4. Analyses of the Feature Scale Range Survey

For each of the 6160 strategies we performed the prediction 10 times using the same 10 CV data. This gives the TSS distributions for the 6160 strategies over different flare classes and CV data. We analyze the result of the survey and find out what factor contributes to the TSS.

First, let $\text{TSS}(s, f, c)$ be the TSS value of the strategy $s$ for flare class $f$, on $c$'th CV data. Here, $s$ belongs to the set of the studied strategies $S$: $s \in S$, $f$ is one of the three classes: $f \in \{X, \geq M, \geq C\}$ and $c \in \mathbb{N}, 1 \leq c \leq 10$.

For a set of real numbers $A$, let $\#A$ be the size of the set. Let $\mu(A)$ and $\sigma(A)$ be the average and the unbiased standard deviation of $A$, respectively, i.e.

$$\mu(A) = \frac{\sum A}{\#A} \tag{4}$$

$$\sigma(A) = \sqrt{\frac{\sum (x - \mu(A))^2}{\#A - 1}} \tag{5}$$

We define the average and the standard deviation of the TSS for each strategy, and for each CV data, as follows:

$$\mu_{f,c} = \mu\{\text{TSS}(s, f, c) \mid s \in S\} \tag{6}$$

$$\sigma_{f,c} = \sigma\{\text{TSS}(s, f, c) \mid s \in S\} \tag{7}$$

$$\mu_{s,f} = \mu\{\text{TSS}(s, f, c) \mid 1 \leq c \leq 10\} \tag{8}$$

$$\sigma_{s,f} = \sigma\{\text{TSS}(s, f, c) \mid 1 \leq c \leq 10\} \tag{9}$$

The $\mu_{f,c}$ and $\sigma_{f,c}$ for every $f$ and $c$ is shown in Figure 6. The figure shows that $\mu_{f,c}$ is distributed in range wider than largest $\sigma_{f,c}$. That is, the choice of CV data have more effect on TSS than choice of the prediction strategy have, despite of the effort to homogenize the prediction difficulty of the CV data as described in §4.3. This means that, if we compare the strategies by $\mu_{s,f} \pm \sigma_{s,f}$, most strategies are within the error bars of each other and we cannot determine the best strategy.

Therefore, in order to determine what better strategies are, we need to subtract the systematic error due to the CV data set dependence. To do this, we calculate the standard scores $z(s, f, c)$, or the z-scores, of the TSS of the strategies. Then we define the average

$\mu^z_{s,f}$ and the standard deviation $\sigma^z_{s,f}$ of the z-scores, as follows:

$$z(s, f, c) = \frac{\text{TSS}(s, f, c) - \mu_{f,c}}{\sigma_{f,c}}, \tag{10}$$

$$\mu^z_{s,f} = \mu\{z(s, f, c) \mid 1 \leq c \leq 10\}, \tag{11}$$

$$\sigma^z_{s,f} = \sigma\{z(s, f, c) \mid 1 \leq c \leq 10\}, \tag{12}$$

Here, $\mu^z_{s,f}$ and $\sigma^z_{s,f}$ are indicators of how a strategy $s$ compares to other strategies at prediction of class $f$. If the $s$ is better than the average of all strategies in $S$, consistently for any CV data set $c$, then $\mu^z_{s,f}$ should be significantly greater than $\sigma^z_{s,f}$.

In Figure 7, we show $\mu^z_{s,f}$ and $\sigma^z_{s,f}$ as functions of strategy $s$, using cross shapes. The horizontal bar marks the range of the feature size for strategy $s$ in terms of $r_l$ and $r_u$. The vertical bar shows the $\sigma^z_{s,f}$ for that strategy. Note that for each feature size range, only one strategy with the best $\mu^z_{s,f}$ is plotted.

As shown in Figure 7, we have found many strategies whose TSS value distributions are significantly better than the average. Note that the line $z = 0$ in Figure 7 indicates a strategy with the average skill score.

Table 4 shows the three strategies that have the best $\mu^z_{s,f}$ for the three classes of events. From Table 4 we can say that these strategies of predicting X, $\geq$M and $\geq$C class flares are better than the average by $2.3\sigma$, $2.1\sigma$, $3.8\sigma$ confidence levels, respectively. Table 5 shows the contingency tables for the best individual predictions for the three classes, together with their TSS values.

The results (Figure 7, Table 4) contain numbers of suggestions, of how we can design good solar flare prediction strategies. To begin with, the best predictors of X-class and $\geq$M-class flares use wavelet inputs with horizontal wavelengths 2% to 7% of the solar diameter, and only those with the largest vertical wavelength. In other words, they are making predictions based on solar images shrinked to several tens of pixels in their widths and only one pixel in the heights.

One possible interpretation of this result is that the shrinking of the images removes any small-scale information and leaves only the most important information — the horizontal difference in the magnetic field, at the scale of the largest active regions. The experimental results suggest that those features contain the essential information for the prediction of the X-class and ≥M-class flares.

In contrast with X-class and ≥M-class flares predictors, the best strategy for ≥C-class flares uses the small-scale information, of $0.002 \leq r_h \leq 0.02$ and $0.002 \leq r_v \leq 0.004$. This again suggests that those scale contains the essential information for ≥C-class flares prediction.

Finally, the combination of standard wavelet transformations with Haar bases resulted in the best predictions for all the classes, compared to the other three combination (c.f. Table 2) that involves either non-standard wavlets or the $\beta$-spline bases.

We have little idea why the prediction of ≥M-class flars have the lowest TSS; generally one would expect a trend from X-M-C or vice versa. The tendency may suggest that our feature space captures most active Sun (with X-class flares) and most quiet Sun (without even C-class flares) relatively well, but not the intermediate states. It would be an interesting feature study to continuously vary the flare X-ray flux threshold from $10^{-6}$Wm$^{-2}$ to $10^{-4}$Wm$^{-2}$ and test how the TSS changes as the result.

## 4.5. Robustness of Prediction Against Input Data Noise

We experimented how TSS of the predictors change as the noise in the input observational data increase. We take the three strategies with the best TSS for the three classes (Table 4), and measured their TSS as we increase the amplitude noise or the time noise. While we measure the amplitude noise and the time noise, we fix the CV data set to [0], in order to study the effect of the introduced noise in isolation.

**Amplitude noise:** The input data with amplitude noise $A_n$ is created from the original data, by multiplying every real number with $\exp(r)$ where $r$ is a random number from uniform distribution of range $(-A_n, A_n)$. That is, the amplitude noise $A_n$ is a non-dimensional value that indicates the ratio of the amplitude of noise over signal. For example, $A_n = 0.1$ means an approximately 10% fluctuation; a signal of $1.0 \times 10^{-4}$Wm$^{-2}$ gets randomly mapped to quantity between $9.048 \times 10^{-4}$Wm$^{-2}$ and $1.105 \times 10^{-4}$Wm$^{-2}$.

**Time noise:** The input data with time noise of timescale $\tau_n$ is created from the original data, by randomly shuffling each time series independently. The shuffle is constrained so that every point in time is moved randomly but no more than $\tau_n$ from the original position. The shuffle is implemented by the following algorithm:

1. Assign each timepoint with a key number. The key number for `i`'th point is $i + r$, where $r$ is a random number from uniform range $0 \leq 1 + \tau_n$.

2. Sort the timepoints in the ascending order of the key number.

Figures 8 and 9 shows the TSS as functions of $A_n$ and $\tau_n$, respectively. The error bars in the figures represent the distribution over the introduced noise.

Figure 8 shows that the TSS values begin to decrease at around $A_n = 0.01$, and decrease rapidly when $A_n > 0.1$. The decrease of TSS is quite monotonic. We conclude that in order to maintain UFCORIN's best performance, the amplitude noise $A_n$ must be kept less than $3 \times 10^{-3}$.

On the other hand, Figure 9 shows that the decrease of TSS is less than 0.01 for $\tau_n < 10$hr, and less than 0.05 for even $\tau_n < 100$hr. The TSS value for prediction of $\geq$ M-class events even increase for $10 < \tau_n < 100$hr. Note that by construction of the time-shuffle algorithm, data is moved both in the future and in the past. Thus, in time noise experiment, "predictions" are made using some knowledge of the future; but the predictors are not merited by these knowledge of the future. These experimental results suggest that UFCORIN is predicting 24-hour GOES X-ray maximum from rather long-term solar features whose timescale is as large as 100 hours.

### 4.6. Comparison with the Related Works

In this section, we compare our results with previous studies, namely those by *Song et al.* [2009], *Bloomfield et al.* [2012], *Ahmed et al.* [2013], and *Bobra and Couvidat* [2014]. We do not compare the TSS values of our and their studies; since the studies differ in the event population they use and how they formulate the forecast, the direct comparison of TSS values has little meaning.

All of the above works study prediction for 24 hours in the future. *Song et al.* [2009] is one of the first studies that systematically predicted X, M, and C class flares. *Bloomfield et al.* [2012] is a good review of flare prediction studies so far, and they proposed the use of TSS as the measure of comparing flare prediction studies. *Ahmed et al.* [2013] uses Solar Monitor Active Region Tracker (SMART) data [*Higgins et al.*, 2011], an automatically-tracked dataset of active regions. They uses the set of 21 designed features as inputs, and applies Cascade Correlation Neural Network (CCNN) to make *operational* prediction, that is, to make prediction for any active region in a given time period. *Bobra and Couvidat* [2014] also uses SHARP data, and present tens of carefully designed feature vectors that are derived from SDO/HMI images, and utilize machine learning technique to construct predictors. They also perform an exhaustive search for the best combination of the feature vector.

There are differences in periods from which training/test data are taken, and differences in what to predict, which introduce uncertainty in the TSS values comparison. To begin with, all the other studies predict flares for active regions (ARs) while we predict flares for the Sun as a whole. *Song et al.* [2009] makes imbalanced selection of X-, M-, and C-class flare events. In *Bobra and Couvidat* [2014], the prediction start time for flaring active regions are set to be exactly 24 hours prior to the flare peak time, while in *Bloomfield et al.* [2012], the prediction start time is the beginning of the day. The latter approach makes each 24-hour prediction statistically independent. On the contrary, since we make 24-hour future prediction with 1 hour cadence, adjacent predictions are not statistically independent in our problem settings. *Ahmed et al.* [2013] may suffer from the same issue since they make 24-hour prediction over 96-minute-cadence magnetic-feature observations.

Even if the event population for all the studies had been the same, we still cannot compare the prediction studies using TSS values. This is because as seen in Figure 6, TSS values fluctuates by changing the cross-validation (CV) data set. The fluctuation is as large as 0.1-0.2 for our same set of prediction strategies. This size of fluctioation is comparable to differences of TSS values reported in different studies, and is 5% of the TSS value range ($-1 \leq \text{TSS} \leq 1$) . This implies that the ranking of different predictors will be easily overturned, when we randomly shuffle the training data and test data while using the same event population. Thus, ranking of different predictors using single CV data set is not stable.

In theory, TSS is not affected by the change of the numbers of positive and negative observations included in the data set [*Bloomfield et al.*, 2012]. Therefore, random permutation of the training and test data, as we have done in this paper, should not affect the TSS values. In practice, however, we saw TSS values fluctuate as we change the CV data set, even when we keep the the difference in numbers of positive and negative observations less than 10% (Figure 6). We point out the two possible causes of the fluctuation: (1) our solar observation data set is never large enough to reflect the "true" distribution of the state of the Sun, and (2) TSS does not correct for the changes of ratio of different kinds of positive events in the dataset, for which some predictors might be good at while other predictors are bad at prediction. Within our study, the source of TSS fluctuation was random permutation of the training and test data. This alone have caused 0.1-0.2 fluctuation in the TSS values.

We emphasize that in order to make viable comparison based on TSS, it is at least necessary to compare different prediction studies on the same CV data set. It is also necessary to use TSS *distribution* over multiple CV data sets, not single TSS value of one CV data set. We can modify our experimental procedure to use per-AR time series as the input, so that we can make per-AR prediction, which makes the comparison between UFCORIN's prediction to the previous studies more viable. We have yet to agree on which per-AR data set we compare. Since in this paper we focus on continuous and full-Sun flare predictions, such per-AR experiment is beyond the scope of the paper and will be a subject of future work.

The global effort to establish flare prediction comparisons is still ongoing. The comparison of flare prediction studies should be based on non-filtered samples, and ideally, on real-time prediction. Efforts are ongoing to establish such form of comparison (c.f. http://ccmc.gsfc.nasa.gov/challenges/flare.php), and automated, continuously-operating flare predictions as achieved by our method are suitable for such form of competition.

## 5. Conclusion and Discussions

Practical flare forecasts should provide predictions 24 hours, 365 days. Therefore, it is important to develop flare prediction methods that operate in fully automated, continuous manner. We established such a fully automated solar flare prediction system. The TSS of our prediction is on the same level as previously reported solar flare prediction studies.

UFCORIN opens extended research in many directions. The first direction is the real-time prediction. Our study so far was on simulated prediction of the past time period (2011-2012). The automation of the real-time prediction is another challenge. We are currently developing the programs required for the automated real-time prediction, such as data-retrieval programs from the open solar databases e.g. *Space Weather Prediction Center Website* [2014b].

UFCORIN's generalized design made it much easier to survey vast solar flare prediction strategies. We are planning to add more input data such as solar images in various wavelength and study the effect of the input data on the predicition skills. We can also test different preprocessing and regression methods.

However, as we increase the variety of the input data, the chance of the missing data will also increase. Missing data arise of various reasons, such as the solar eclipse, or the maintainance, mulfunction or the discontinuation of an observation apparatus. How to deal with missing data is one of the open questions in machine learning [*Ghahramani and Jordan*, 1994; *Smola et al.*, 2005; *Marlin*, 2008; *Jerez et al.*, 2010].

UFCORIN's robustness against time noise (c.f. §4.5) may provide a solution for missing data. In case of missing data event, we can randomly assign the available data from within ±100 hr to the missing points. UFCORIN's prediction skill is not affected by this kind of operation (c.f. Figure 9).

Yet another important use of UFCORIN is to search for the ways to shrink the input data, without worsening the TSS. Importance of shrinking the data has already been demonstrated in this paper. If we find a input data set of smaller size that achieves relatively high TSS, those channels in the input data set are likely to contain physical information of the solar flare trigger. Thus, we can use UFCORIN to search for the empirical triggers of the solar flares.

However, we would like to acknowledge that as we increase the number of different strategies without increasing the dataset , we risk statistical flukes — in other words, attaining high scoring combinations by chance. With 6,160 different strategies we have studied, we already risk statistical flukes, and investigating how much of the variation is by chance is an interesting future works. Again, the real time prediction experiment will distinguish truly correct predictions from correct predictions by chance.

It is easy to configure UFCORIN to predict quantities other than solar X-ray flux, such as mass and speed of coronal mass ejections, total flux and power indices of the solar energetic particles, or even the Dst index. UFCORIN has wide application in the space weather.

Although we have used only TSS to evaluate the prediction result in this study, UFCORIN can be used to optimize prediction values other than TSS. There are many human activities that are affected by space weather [*National Research Council*, 2008]. Satellite operating companies want to minimize flare damage. Airline operators need to avoid noise in HF band. Operators of solar observation satellites such as Hinode and IRIS want to point the telescope at the sunspots where large flares are expected. Different agencies have different benefits from predictions. Their losses from false predictions also differ. Once the quantitative values of four prediction scenarios (true positive, true negative, false positive, and false negative) are given, UFCORIN can be used to provide customized space weather prediction for different agencies so that their benefits are maximized.

**Acknowledgments.**

This work was supported by a Grant-in-Aid from the Ministry of Education, Culture, Sports, Science and Technology of Japan (No. 25287039). Part of this research used the computational resources supported by the RIKEN Advanced Institute for Computational Science(AICS). We thank Ayumi Asai and Shinsuke Takasao for careful reading of the draft version of this paper. We thank Josh Barnes for the thorough check of the English of the paper. And we thank the annonymous referees for the careful reading of our paper, which resulted in substantial improvement of the paper.

This is a joint research project of Kyoto University and BroadBand Tower, Inc. We thank Hiroshi Fujiwara, the CEO of BroadBand Tower, Inc. for his continuous support and encouragement that made this research possible in the first place. We thank NASA SDO/HMI team and the GOES team for the data used in this study.

**Appendix A: Details of the Implementation**

In this appendix, we describe details of UFCORIN for the convenience of people who want to use UFCORIN.

**A1. Timeline Data File Format**

A Timeline Data File (TLDF) format is a simple text-based format for time-dependent data. A TLDF consists of lines of whitespace separated strings. A '#' and every characters that follow it in a line are comments. Lines are separated to words by one or more sequence of whitespace characters. The first sequence of non-whitespace characters in a line are the column 1, and so on.

In UFCORIN, each time series data is specified by a pair of a filename and a schema. The file contains the actual data in TLDF format, and the schema specifies how to read the file.

An example of a TLDF file and a schema specification of that file are shown in Figure 10. As is shown, a schema consists of the four fields. The two fields `colT` and `colX` specify the column indices of the timebin column and the actual data column. If the `isLog` field is `true`, the natural logarithms of the data is used in regression, rather than the raw data. Finally, `scaling` specifies the scaling factor $s$ of the data. In other words, the schema defines the relation between the raw data $x'_i$ and the feature vector data $x_i$ that are used by the regressor. The relation between $x_i$ and $x'_i$ are:

$$x_i = sx'_i, \tag{A1}$$

if `isLog : false`, and

$$x_i = s\log(x'_i), \tag{A2}$$

if `isLog : true`.

Timebin is an integer that specifies a certain span of time. Timebin is the index of the time series, and it can be any number that is an increasing function of time. In our current convention, each timebin is one hour long, and timebin 0 is the first one hour in the year 2011, UTC. In other words, timebin 0 is the set $\{t \mid 2011/01/01\text{T}00:00:00\text{Z} \leq t < 2011/01/01\text{T}01:00:00\text{Z}\}$.

## A2. Strategy File Format

Testing a new prediction strategy is very easy in UFCORIN. The only thing the user needs to do is to write the *strategy file*. The syntax of the strategy file is based on YAML, a human-readable data format proposed by *Ben-Kiki et al.* [2009] and is widely used. We describe the contents of the strategy file with an example (Figure 11 .)

The first field, `spaceWeatherLibVersion`, specifies the version of the strategy file. As we developed UFCORIN we added more features, and sometimes we had to update the format definition of the strategy file. In such cases we increase the strategy file format version. Different versions of the formats are mutually incompatible.

The next field, `crossValidationStrategy`, specifies the time interval for cross validation. The current UFCORIN supports `Weekly`, `Monthly`, and `Yearly` cross validation. The input and output data are separated into chunks of the equal time intervals. The chunks are indexed, starting from zero. Then, chunks with even indices are used as training data while chunks with odd indices are used as test data. For any cross validation strategy, UFROCIN also supports its `Negate`, where the training data and the test data are swapped.

The next two fields, `predictionTargetFile` and `predictionTargetSchema`, specifies the output time series. This is the time series to be predicted. See section A1 for the details of the time series data format.

The next field, `regressorUsed`, specifies the name of the regression algorithm to be used. The parameters of the regression algorithm are also specified here. The regression algorithm is the core of the prediction, since its role is to construct the model function $y(\vec{x})$. Two regression engines are integrated into UFCORIN and are available at the moment: one is LibSVM, a support vector machine (SVM) library by *Chang and Lin* [2011]. The other is a simple, handwritten liner regression algorithm. They are specified by tags `LibSVMRegressor` and `LinearRegressor`, respectively.

The SVM regressor of LibSVM has multiple parameters such as $C$ (`Cost`) and $\gamma$ (`Gamma`) that affects the performance of the regressor. These parameters can be specified directly in the strategy file. Furthermore, when the value of `AutomationLevel` is set to positive integers, UFCORIN automatically optimizes the machine learning parameters using Covariance Matrix Adaptation Evolution Strategy (CMA-ES) [*Hansen and Ostermeier*, 1996, 2001].

`AutomationLevel`= 1 instructs UFCORIN to maximize the sum of TSS for the three classes of events X, ≥M, and ≥C. `AutomationLevel`= 2, 3, 4 instructs UFCORIN to maximize the TSS for X, ≥M, and ≥C class of events, respectively.

The final field of the strategy file specifies the input data to be used to construct the prediction model. This `featureSchemaPackUsed` field is divided into two subfields. In the first field, `SchemaDefinitions`, the users can define schemas they want to use for the input files, and assign each schema definition to a label. In the second field, `FilenamePairs`, the users specify the input data as pairs of a schema labels and a data file name. We adopt this two-part format because the input file lists tend to contain hundreds of files in the real usecases. In such cases, a large number of files come from the same data source and share exactly the same schema. Therefore, pre-defining the schema labels and using the label-filename pairs make the input files specification more readable and more space-efficient, compared to directly providing one schema per one filename.

**A3. Computer Resources**

From August 2013 to September 2014, UFCORIN has been hosted on servers owned by BroadBand Tower, Inc. There were ten dedicated server nodes for UFCORIN. Each node was equipped with Intel Xeon (4 core) CPU and 16-Gigabyte RAM. Hadoop, an open-source map & reduce framework, and Hadoop Distributed File System (HDFS) were installed on the system. Of the ten nodes, two were master nodes, that acts as HDFS name nodes and Hadoop job tracker nodes. The other eight nodes were HDFS data nodes and Hadoop task nodes. The total capacity of HDFS was 50TB.

Since September 2014, the system has been migrated to Amazon Web Services (AWS). The computer resource demand of UFCORIN fluctuate because it is still under development. Therefore virtualized, on-demand computer resources provided by AWS allows us more effective use of our research budget.

**Appendix B: Programs and Data Availability**

In compliance with AGU's Data Policy, we are eager to provide access to the computer programs and the data we have used in this research. The source code for UFCORIN is published at `https://github.com/nushio3/UFCORIN`. The data is hosted at Amazon S3 (Simple Storage Service); please contact authors for access to the data. The data was originally obtained from SDO/HMI and GOES websites, via URLs `http://satdat.ngdc.noaa.gov/sem/goes/data/` and `http://jsoc.stanford.edu/`.

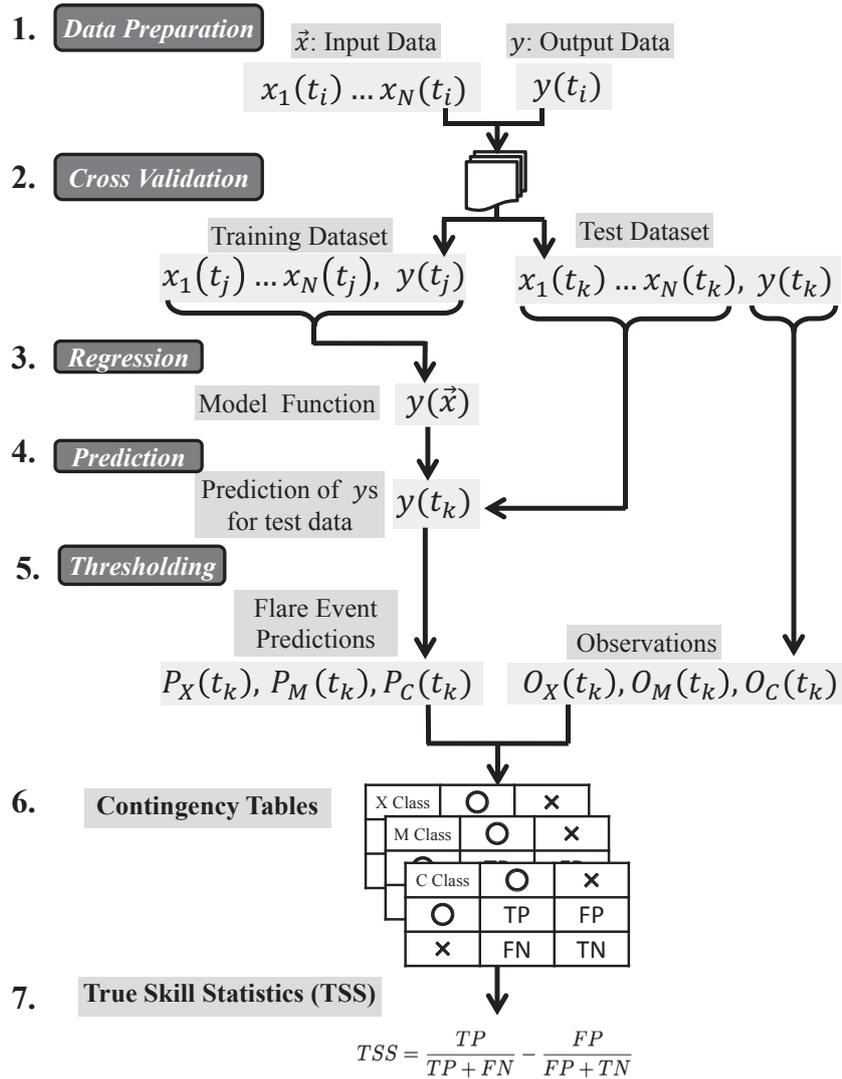

**Figure 1.**  The flowchart of the forecast procedure of UFCORIN.

| event | Observed | Not observed |
|---|---|---|
| Predicted "Yes" | $TP$ | $FP$ |
| Predicted "No" | $FN$ | $TN$ |

**Table 1.** A contingency table consists of four numbers. They are counts for true positive (TP), true negative (TN), false positive (FP), and false negative (FN) cases, respectively.

| Basis label | description |
|---|---|
| `haar-S` | centered Haar wavelet, standard form |
| `haar-N` | centered Haar wavelet, non-standard form |
| `bspl-S` | centered B-spline wavelet of order (3,1), standard form |
| `bspl-N` | centered B-spline wavelet of order (3,1), non-standard form |

**Table 2.** The list of wavelet bases used in this study, and their label.

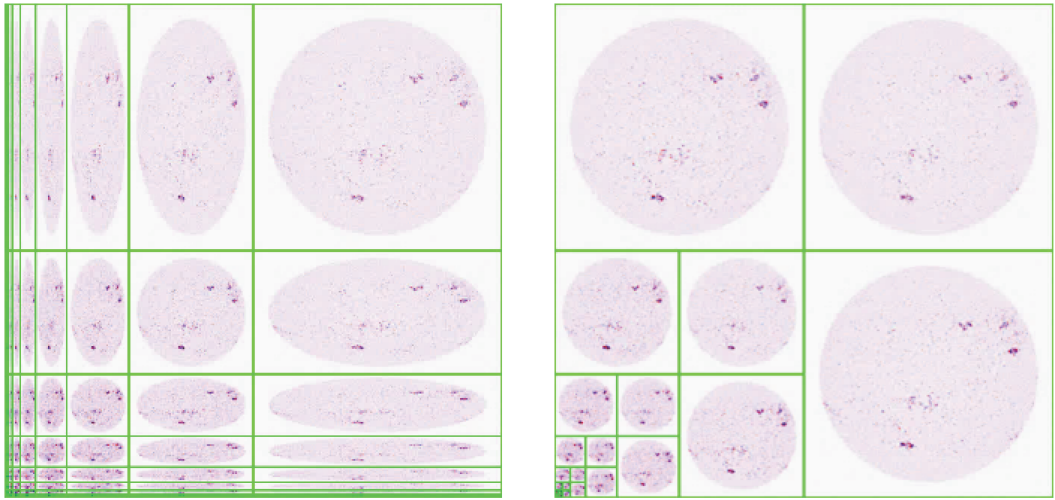

**Figure 2.** Visual demonstration of wavelet transformations used in this study. (row 1) An image of disk filled with random real numbers of range $(-1, 1)$, transformed by centered Haar wavelets `haar-S` and `haar-N`. (row 2) Projection of some of the `haar-S` bases onto the image space. (row 3) Projection of `haar-N` bases. (row 4) Projection of some `bspl-N` bases onto the image space.

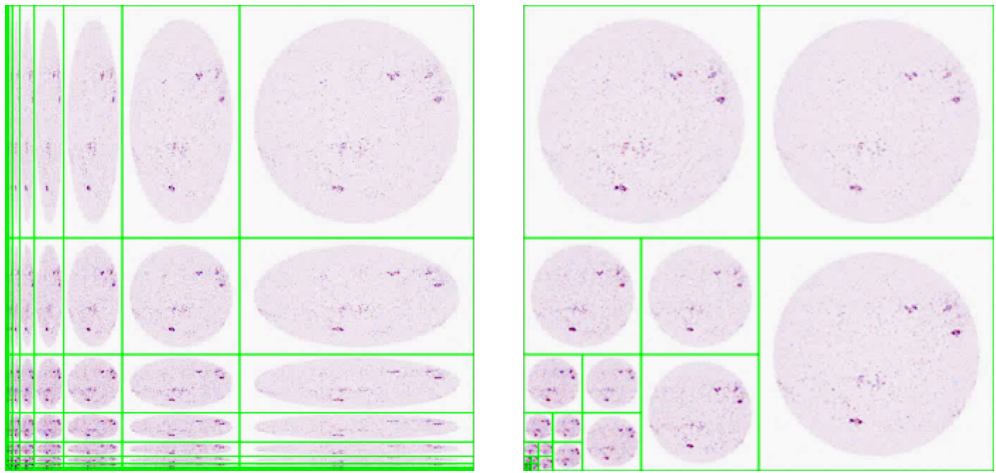

**Figure 3.** The result of applying standard (left) and non-standard (right) Haar-wavelet transformation to the SDO/HMI image. Wavelet space regions that correspond to different horizontal/vertical scales are marked by green rectangles.

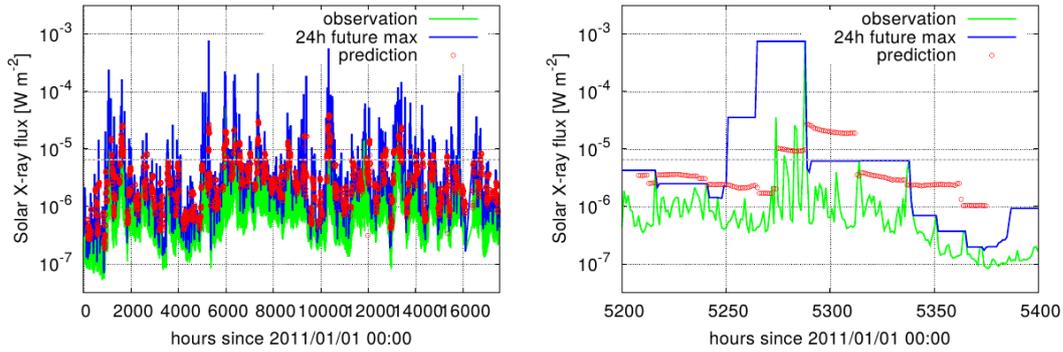

**Figure 4.** Green curves: soft X-ray flux (1-8 Å) observed by GOES satellite; Blue curves: the maximum of the X-ray flux for the next 24 hours in the future ($y_{\max,24\mathrm{hr}}(t)$); Red points: predicted values of $y'_{\max,24\mathrm{hr}}(t_k)$, by UFCORIN; Thick dashed gray line: the threshold for predicting X-class positive events. The left graph shows the entire 2-year period used in this study, while the right graph shows a 200-hour period.

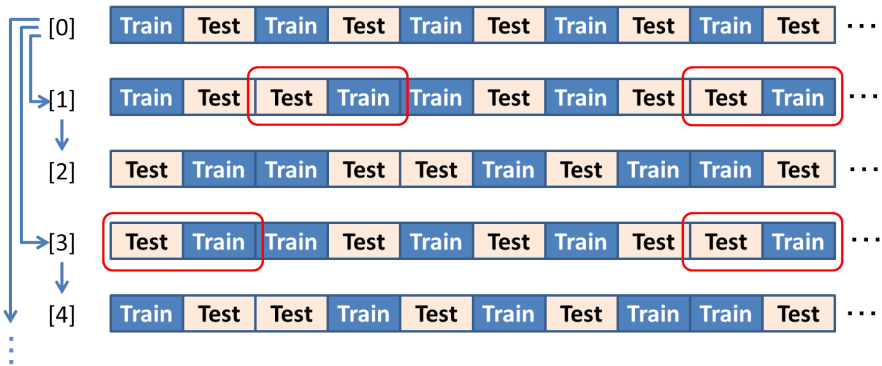

**Figure 5.** Our method for construction of the cross-validation data. Each rectangle marked 'Train' or 'Test' corresponds to a time segment of one week. Starting from the alternating assignment ([0]), we create the odd-numbered data ([1], [3], ...) by randomly swapping the role of the adjacent weeks and screening by the ratio of flaring/non-flaring events. The even numbered data ([2], [4], ...) are the negatives of their predecessors.

| CV | X class | | ≥M class | | ≥C class | |
|---|---|---|---|---|---|---|
| [1] | 167 | 172 | 1706 | 1633 | 6447 | 6498 |
| [3] | 172 | 167 | 1721 | 1618 | 6601 | 6344 |
| [5] | 170 | 169 | 1733 | 1606 | 6537 | 6408 |
| [7] | 171 | 168 | 1627 | 1712 | 6567 | 6378 |
| [9] | 177 | 162 | 1667 | 1672 | 6242 | 6703 |

**Table 3.** The number of positive events for each class in the training (left) and test (right) data for CV data sets [1], [3], [5], [7], and [9]. The CV data sets [2], [4], [6], [8], and [10] are created by swapping the training and test data from CV data sets [1], [3], [5], [7], and [9], respectively.

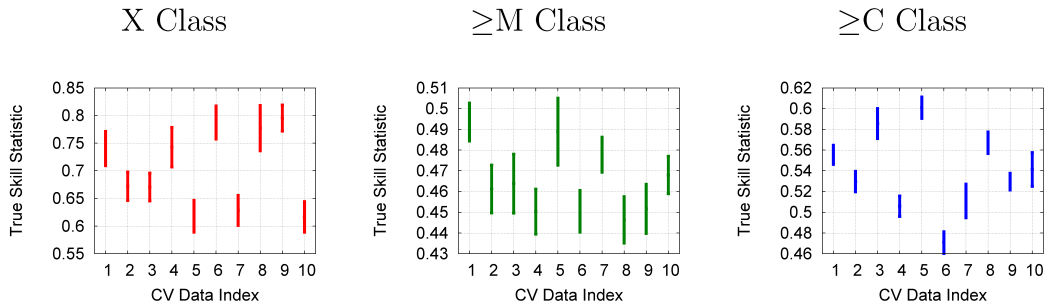

**Figure 6.** The distribution of TSS values for every CV data, every flare classes ($\mu_{f,c}$, $\sigma_{f,c}$ ).

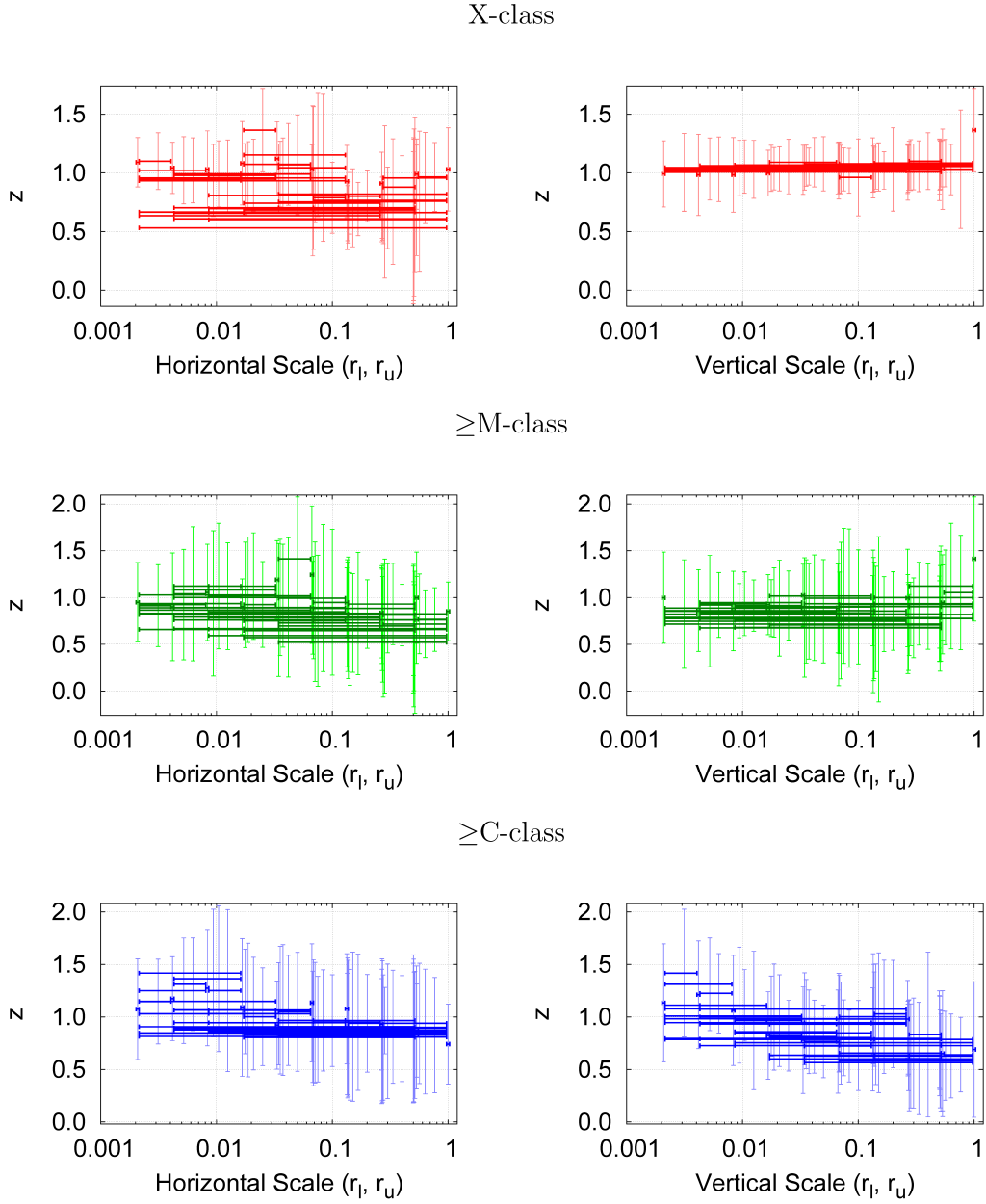

**Figure 7.** The TSS distribution of the prediction strategies as we limit the range of the wavelet features size in the input set. Each cross shape represents a strategy, where the horizontal bar is its range of the wavelet bases and the vertical bar represents the mean and the standard deviation of its TSS.

X - 44    MURANUSHI ET AL.: SOLAR FLARE FORECAST IN GOES X-RAY FLUX| | $w$ | $r_{hl}$ | $r_{hu}$ | $r_{vl}$ | $r_{vu}$ | $\mu_{s,f} \pm \sigma_{s,f}$ | $\mu^z_{s,f} \pm \sigma^z_{s,f}$ |
|---|---|---|---|---|---|---|---|
| $s_X$ | haar-S | 0.02 | 0.03 | 1. | 1. | $0.745 \pm 0.072$ | $+1.42 \pm 0.61$ |
| $s_{\geq M}$ | haar-S | 0.03 | 0.07 | 1. | 1. | $0.481 \pm 0.017$ | $+1.41 \pm 0.66$ |
| $s_{\geq C}$ | haar-S | 0.002 | 0.02 | 0.002 | 0.004 | $0.557 \pm 0.043$ | $+1.36 \pm 0.36$ |

**Table 4.** The strategies with the best $\mu_{s,f}$ for each $f$.

| X class flare | Observed | Not observed |
|---|---|---|
| Predicted "Flare" | 150 | 749 |
| Predicted "No Flare" | 10 | 7440 |
| TSS = 0.846 | | |
| ≥M class flare | Observed | Not observed |
| Predicted "Flare" | 1336 | 2266 |
| Predicted "No Flare" | 238 | 4571 |
| TSS = 0.517 | | |
| ≥C class flare | Observed | Not observed |
| Predicted "Flare" | 5021 | 373 |
| Predicted "No Flare" | 1237 | 1780 |
| TSS = 0.629 | | |

**Table 5.** Contingency tables for X, ≥M, and ≥C class flare prediction by UFCORIN.

D R A F T        October 31, 2015, 3:00pm        D R A F T

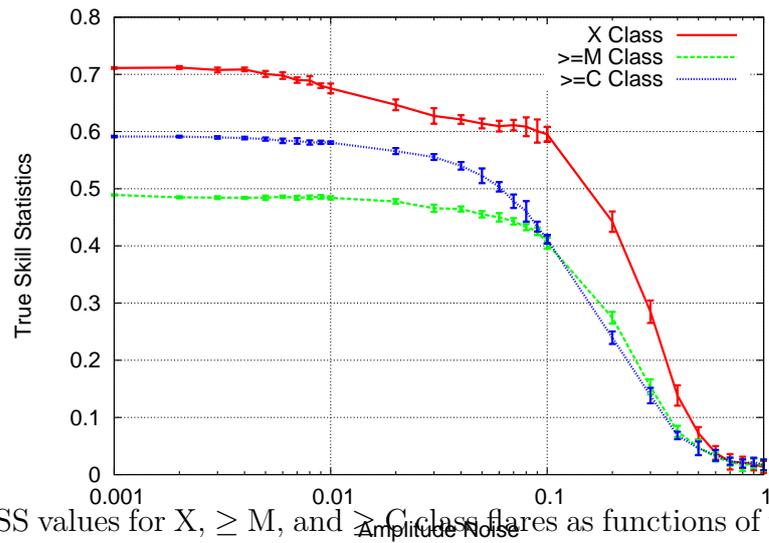

**Figure 8.** The TSS values for X, $\geq$ M, and $\geq$ C class flares as functions of the amplitude noise $A_n$.

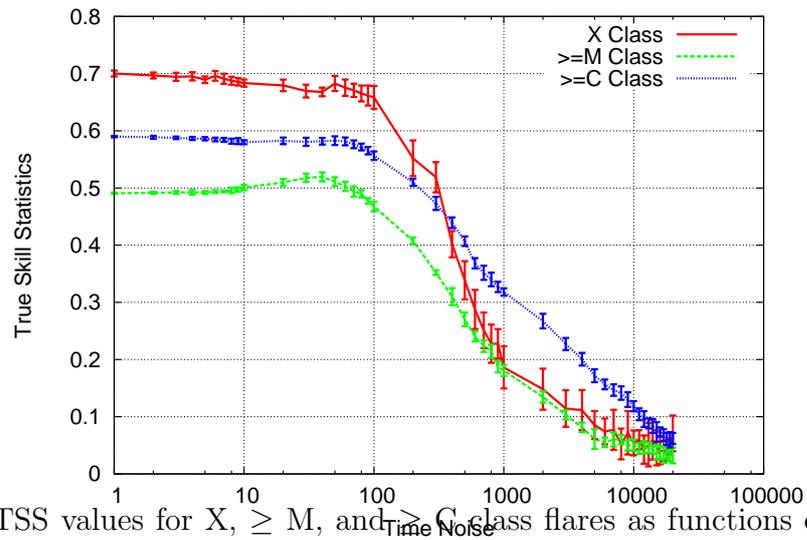

**Figure 9.** The TSS values for X, $\geq$ M, and $\geq$ C class flares as functions of the time noise $\tau_n$. The horizontal axis is in the units of hours.

```
/user/nushio/wavelet-features/bsplC-301-N-0000-0000.txt
```
```
...
2011-02-21 10 1234 7.063820800780941e2 4.989756430554549e5
2011-02-21 11 1235 7.666781616210947e2 5.877954035067013e5
2011-02-21 12 1236 6.391829833984789e2 4.0855488626618014e5
2011-02-21 13 1237 6.498704833984486e2 4.223316451925333e5
2011-02-21 14 1238 8.193226928710915e2 6.712896750535369e5
...
```

```
/user/nushio/strategies/sample.yml
```
```
...
{ colT: 3, colX: 5, isLog: true, scaling: 1.0}
...
```

**Figure 10.** An example of a pair of a raw TLDF file and a schema specification of that file.

```
/user/nushio/strategies/sample.yml

spaceWeatherLibVersion: version 1.1

crossValidationStrategy: {tag: CVWeekly, contents: []}

predictionTargetFile: /user/nushio/forecast/forecast-goes-24.txt

predictionTargetSchema: {colT: 2, colX: 5, isLog: true, scaling: 1.0}

regressorUsed:

  - tag: LibSVMRegressor

    contents: { Cost: 100.0, Nu: 0.5, Epsilon: 1.0e-3,

                Gamma: null, KernelType: 2, Type: 3, AutomationLevel: 0}

featureSchemaPackUsed:

  SchemaDefinitions:

    f25Log: { colT: 2, colX: 5, isLog: true, scaling: 1.0}

    f35Log: { colT: 3, colX: 5, isLog: true, scaling: 1.0}

  FilenamePairs:

  - [f35Log, /user/shibayama/sdo/hmi/hmi_totalflux.txt]

  - [f35Log, /user/nushio/wavelet-features/bsplC-301-N-0000-0000.txt]

  - [f25Log, /user/nushio/forecast/backcast-goes-24.txt]
```

**Figure 11.** A sample strategy file.